\documentclass{webofc}
\usepackage[varg]{txfonts}   
\usepackage{hyperref}
\usepackage{url}
\hypersetup{colorlinks=true,citecolor=blue,urlcolor=blue,linkcolor=blue}
\begin{document}
\title{Probing light nuclei production mechanism by measuring nucleus production in and out of jets}
%
%

\author{\firstname{Chiara} \lastname{Pinto}\inst{1}\fnsep\thanks{\email{chiara.pinto@cern.ch}}
        \lastname{for the ALICE Collaboration} 
}

\institute{European Organisation for Nuclear Research (CERN), Geneva, Switzerland
          }

\abstract{The production mechanism of (anti)nuclei in ultrarelativistic hadronic collisions is under debate in the scientific community. Two successful models used for the description of the experimental measurements are the statistical hadronization model and the coalescence approach. In the latter, multi-baryon states are assumed to be formed by the coalescence of baryons that are close in phase-space at kinetic freeze-out. Given the collimated emission of nucleons in jets, the available phase-space is limited. As a result, the production of nuclear states through coalescence in jets is expected to be enhanced compared to production in underlying events. In this contribution, the results for the coalescence parameter $B_2$, which quantifies the formation probability of deuterons by coalescence, measured in and out of jets with the ALICE detector at the Large Hadron Collider, are presented and discussed in the context of the coalescence model.
}
\maketitle
\section{Introduction}
High-energy collisions at the Large Hadron Collider (LHC) provide an ideal environment for investigating the production of light (anti)(hyper)nuclei. At center-of-mass energies ranging from approximately 1 to 13 TeV, these collisions produce nearly equal amounts of matter and antimatter, given the baryochemical potential close to zero ($\mu_{\rm B} \sim 0$). This symmetry enables a detailed study of hadronization processes under conditions similar to those of the early Universe.

Understanding and modeling the mechanism responsible for (anti)nuclei formation is essential for multiple reasons. First, it advances our fundamental knowledge of Quantum Chromodynamics (QCD) in its non-perturbative regime, providing insights into the transition from quark-gluon matter to hadronic states. Second, precise measurements of (anti)nuclei production are critical for astrophysical research, particularly in interpreting cosmic-ray antinuclei fluxes, which could serve as indirect signals of dark matter annihilation or decay~\cite{Donato:1999gy}. 
Among others, two phenomenological models describe nucleosynthesis, namely the statistical hadronization model (SHM) and baryon coalescence. 

The production of nuclei in jets, and its comparison with that in underlying events, offers a unique scenario in which the phase-space available for nucleons is constrained by the jet definition, and the production through coalescence can be tested. 

In this article, recent measurements from the ALICE Collaboration of the nuclear production in jets and in the underlying event, are presented. Experimental results are contextualized with current theoretical frameworks, highlighting advances and challenges in the field.

\section{Production models}
\subsection{Statistical Hadronization Models (SHM)}
Statistical hadronization models assume that hadrons are produced from a thermally equilibrated system described by a temperature ($T_{\mathrm{chem}}$), at which the chemical species are fixed, a correlation volume ($V_{\rm c}$) in which the system is in equilibrium because of conservation of baryon number, charge, and strangeness content, and a baryochemical potential ($\mu_{\rm B}$), which can be fixed based on the antimatter-to-matter ratios. In this framework, particle yields are given by
\begin{equation}
\frac{\rm{d}\it{N}}{\rm{d}\it{y}} \propto \exp\left(-\frac{m}{T_{\mathrm{chem}}}\right),
\end{equation}
which implies a strong sensitivity of nuclei (which have a larger mass, i.e., from $\sim$1 to $\sim$3 GeV/$c^2$, compared to the bulk of particles, i.e., pions, with a mass of $\sim$140 MeV/$c^2$) to the chemical freeze-out temperature. SHM has been very successful in reproducing the integrated yields in heavy-ion collisions and, with adjustments using the canonical ensemble for small systems, in pp and p--Pb collisions \cite{andronic}.

\subsection{Coalescence models}
Coalescence models provide a microscopic description of light (anti)nuclei production by considering the proximity of (anti)nucleons in phase-space, utilizing the Wigner function formalism~\cite{butler,mahlein}. In this approach, nuclei form when constituent nucleons exhibit sufficient overlap of their phase-space distributions with the Wigner density of the bound state. 
The nucleus formation probabilities are quantified by the coalescence parameter $B_{\rm A}$, defined as 
\begin{equation}
B_{\rm A} \propto \frac{N_{\rm A}}{(N_{\rm p})^{\rm A}},
\end{equation}
where $N_{\rm A}$ represents the measured invariant yield of nuclei with mass number $A$, and $N_{\rm p}$ denotes the proton invariant yield. Notably, the transverse momentum of the proton corresponds to that of the nucleus scaled by the mass number ($p_{\rm T}^{ \rm p} = p_{\rm T}^{ \rm A}/A$). 

The comparison between the measured coalescence parameter $B_{\rm A}$~\cite{alice_jhep} and the state-of-the-art model predictions~\cite{mahlein} has shown the importance of both the size of the particle emission region (the so-called source size) and the internal wave function of the nucleus itself, in the formation process. 

To further test the coalescence model, measurements of the deuteron production were carried out within high-density jet environments by the ALICE Collaboration. Such studies reveal significantly enhanced values of $B_{\rm A}$ in jets compared to the corresponding $B_{\rm A}$ in the underlying event (UE), thus validating the coalescence scenario, in which nucleus formation probability increases with local nucleon density in phase-space~\cite{nucleiinjets}.

\section{Experimental setup and analysis techniques}
\subsection{ALICE detector}
The ALICE detector is uniquely suited for the identification of light nuclei, thanks to its low material budget and excellent tracking and particle identification (PID) capabilities. A detailed description of the ALICE subdetectors and their performance can be found in Refs.~\cite{alice_general_2, ALICEperformance} and related references. Among the detectors of the central barrel (the central part of the experiment surrounded by the solenoidal magnet), two are used for the PID of (anti)protons and (anti)deuterons:
\begin{itemize}
    \item Time Projection Chamber (TPC): Provides excellent particle identification through energy loss (d$E/$d$x$) measurements in the low momentum region (below 1 GeV/$c$);
    \item Time-of-Flight (TOF): Used for velocity ($\beta$) measurements at higher momenta \mbox{($>1$ GeV$/c$)}, offering complementary PID capabilities to the TPC.
\end{itemize}

\subsection{Jet and underlying event separation}
For the results shown in this article, events that pass the selection of a leading track with transverse momentum above 5 GeV$/c$ are classified into three regions. The regions are defined depending on the azimuthal angle of each track with respect to the azimuthal angle of the leading track, which is assigned to be zero. The regions are therefore the following:
\begin{itemize}
    \item Toward region: $|\Delta \phi| < 60^\circ$
    \item Transverse region: $60^\circ < |\Delta \phi| < 120^\circ$
    \item Away region: $|\Delta \phi| > 120^\circ$
\end{itemize}
The jet distributions are obtained by subtracting the underlying event contribution (assumed to be dominating the Transverse region) from the Toward region (which instead contains jets in addition to the underlying event). This method allows for a differential study of the coalescence parameter $B_2$ and the deuteron-to-proton (d/p) ratios in the different regions.

\section{Results and discussion}
\subsection{Coalescence parameter measurements}
\begin{figure}[!hbt]
\centering
\begin{minipage}[b]{0.49\textwidth}
    \includegraphics[width=\textwidth]{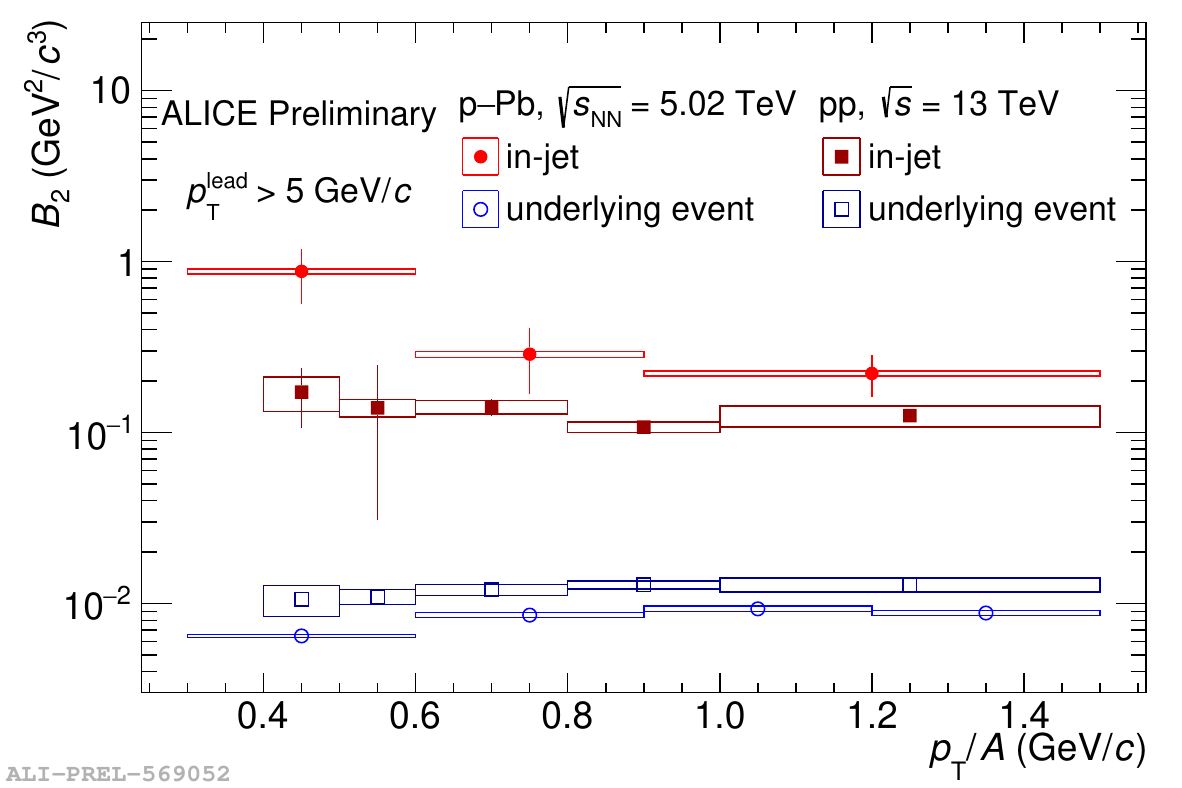}
\end{minipage}   
\hfill
\begin{minipage}[b]{0.49\textwidth}
    \includegraphics[width=\textwidth]{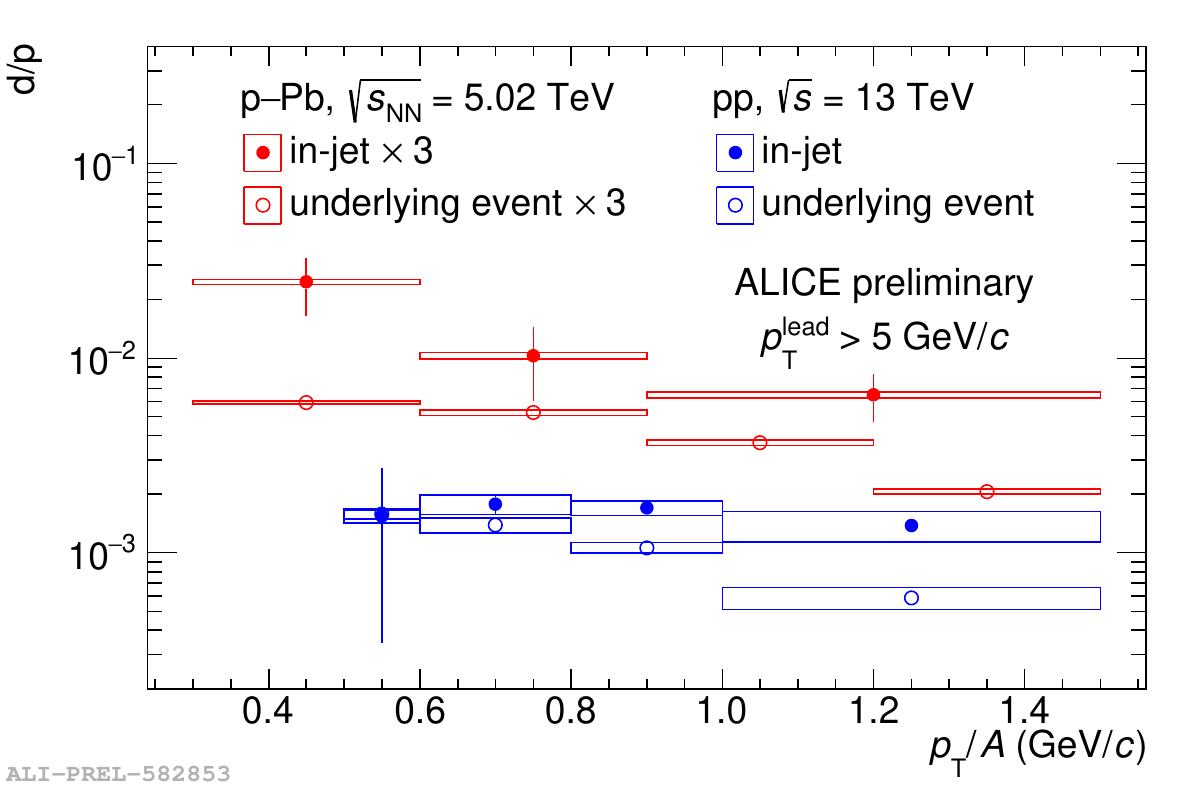}
\end{minipage}
\caption{Coalescence parameters $B_2$ (left panel) and deuteron-to-proton yield ratio (right panel) as a function of $p_{\mathrm T}$/$A$, measured in pp and p--Pb collisions at different energies. Statistical uncertainties are shown as vertical lines whereas boxes represent systematic uncertainties. }
\label{fig:CoalescenceParameters}
\end{figure}

The coalescence parameters measured in jets and UE, in pp collisions at $\sqrt{s} = $13 TeV~\cite{nucleiinjets} and p--Pb collisions at $\sqrt{s_{\rm NN}} = $5.02 TeV, are shown in the left panel of Fig.~\ref{fig:CoalescenceParameters}. Experimental data show a significant enhancement of the coalescence parameter $B_2$ in the jet region compared to the UE, in both pp and p--Pb collisions. This enhancement is interpreted as a direct consequence of the reduced phase-space separation of nucleons in jets with respect to the UE. Furthermore, comparisons between pp and p--Pb collisions show that the in-jet $B_2$ is more enhanced in p--Pb compared to pp collisions, and that the UE $B_2$ is smaller in p--Pb than in pp collisions. The latter measurement can be understood considering that in pp collisions the emission source size is smaller ($\sim1$~fm~\cite{sourcepp}), compared to that in p--Pb ($\sim1.5$~fm~\cite{sourcepPb}), reflecting in a higher $B_2$ for pp collisions in the UE. The former, instead, is still not entirely understood. It might indicate that the particle composition in jets is different between pp and p--Pb, reflecting different abundances of nucleons available for coalescence in the jet depending on the collision system. 
However, these considerations are not yet proven conclusively, and further studies in this direction are needed, also exploiting the larger statistics available with the LHC Run 3 data taking.
Finally, the yield ratios between deuterons and protons, shown in the right panel of Fig.~\ref{fig:CoalescenceParameters}, suggest similar interpretations regarding the different particle composition of jets in pp and p--Pb collisions. 

\subsection{Comparison to coalescence models}
\begin{figure}[!hbt]
\centering
\begin{minipage}[b]{0.49\textwidth}
    \includegraphics[width=\textwidth]{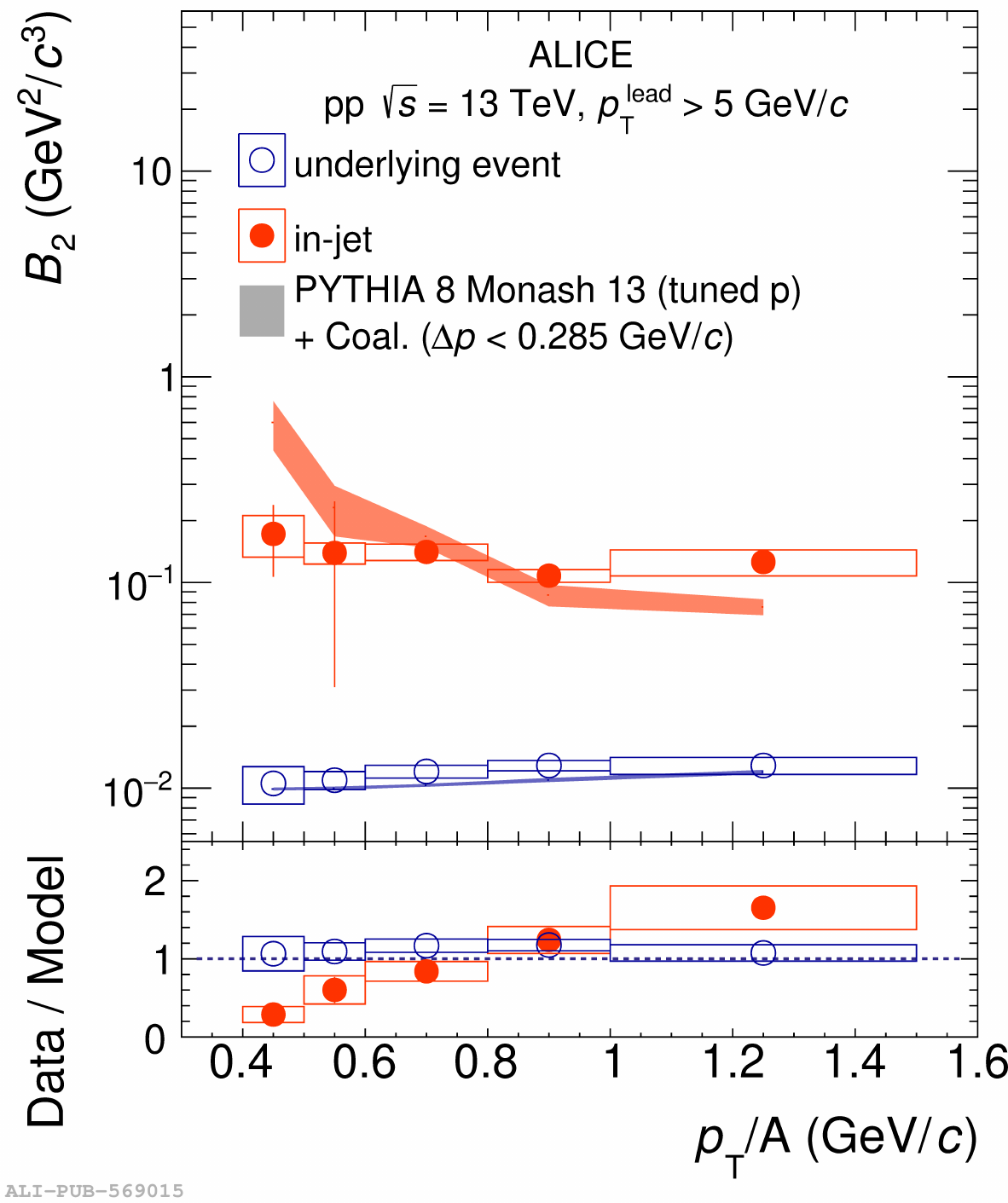}
\end{minipage}   
\hfill
\begin{minipage}[b]{0.49\textwidth}
    \includegraphics[width=\textwidth]{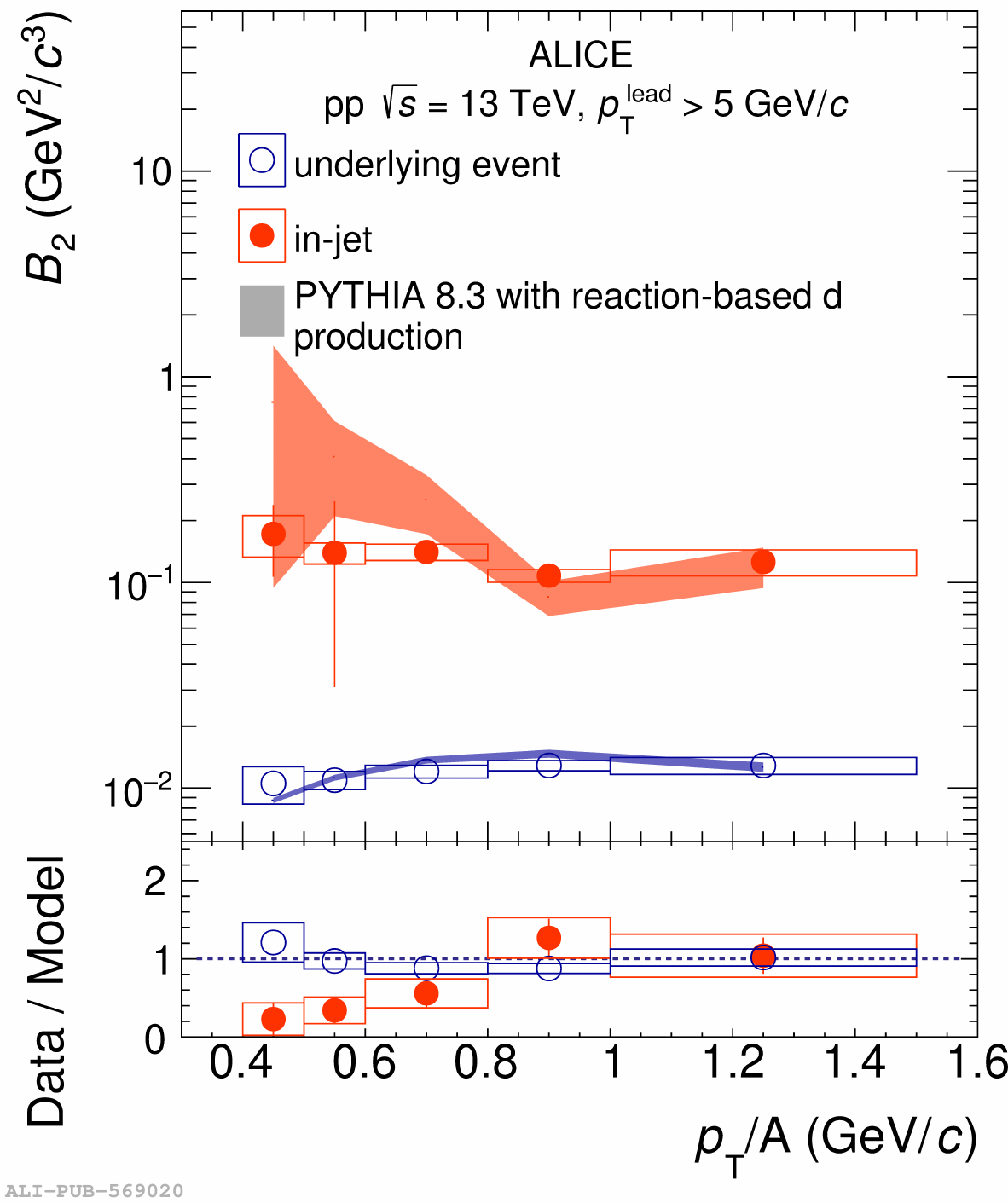}
\end{minipage}
\caption{Coalescence parameters $B_2$ in jets (red) and UE (blue) as a function of $p_{\mathrm T}$/$A$, measured in pp collisions compared to PYTHIA 8 + coal. model (left panel) and PYTHIA 8.3 (right panel). Statistical uncertainties are shown as vertical lines whereas boxes represent systematic uncertainties. }
\label{fig:models}
\end{figure}

The experimental results from pp collisions are compared to the predictions of simple coalescence models using Monte Carlo (MC) simulations (see Fig.~\ref{fig:models}). Namely, two models are developed, the first uses PYTHIA 8~\cite{skands} as an event generator and a spherical coalescence approach for the formation of deuterons (according to which deuterons are formed if protons and neutrons are close in momentum coordinates, with a difference in momentum $\Delta p < p_0 = 0.285$ GeV/$c$). The second model utilizes PYTHIA 8.3~\cite{martin} as an event generator, and in this implementation of the MC, deuterons are produced with energy-dependent parameterizations of nuclear reactions. All the details about the two models are given in Ref.~\cite{nucleiinjets}. Both models reproduce the large gap between the $B_2$ in jets and UE, as well as the trend of $B_2$ as a function of the reduced transverse momentum in the UE, while discrepancies in jet regions indicate that further theoretical refinements are needed. 

\section{Conclusions and future perspectives}
The study of (anti)nuclei production in jets and the underlying event with ALICE is a powerful tool to explore new insights into the nuclear coalescence mechanism. Enhanced coalescence probability in jets with respect to the underlying event is observed in both pp and p--Pb collisions. These results support the hypothesis that the local phase-space density plays a critical role in nucleus formation. These findings have important implications for both nuclear physics and astrophysics, particularly in the context of indirect dark matter searches, as modeling the nucleus formation mechanism is crucial for describing the background of the antinuclear fluxes from cosmic rays. With more data collected during the LHC Run 3, further differential measurements will enable even more stringent tests of current theoretical models. 
Particularly, during the LHC Run 3 data taking, the collected datasets have a three order of magnitude increase over Run 2 data, which will enable more differential studies, including: more refined jet selections using the anti-$k_{\rm T}$ algorithm, studies of (anti)nuclei production as a function of jet radius and multiplicity, exploration of additional light nuclei (e.g., $^3$He and $^4$He) to further test theoretical models.

\end{document}